\documentclass[a4paper,twocolumn,8pt]{revtex4}
\usepackage{graphicx}
\usepackage{fancyhdr}
\usepackage{amsmath,amsfonts,amssymb}

\usepackage{color}
\definecolor{red}{rgb}{1,0,0}
\definecolor{gre}{rgb}{0,1,0}
\definecolor{blu}{rgb}{0,0,1}

\newcommand{\lb}[1]{\label{#1}}
\newcommand{\ff}[1]{(\ref{#1})}

\newcommand{\bark}{\bar{k}}
\newcommand{\barp}{\bar{p}}
\newcommand{\barmu}{\bar{\mu}}
\newcommand{\barn}{\bar{N}}
\newcommand{\hk}{\hskip 0.2truecm}

\newcommand{\barN}{\bar{N}}
\newcommand{\baralpha}{\bar{\alpha}}
\newcommand{\barnu}{\bar{\nu}}
\newcommand{\barpi}{\bar{\pi}}

\newcommand{\K}[1]{\mathbb{K} \left[#1 \right]}

\begin{document}

\title{Gauge-invariance in Loop Quantum Cosmology : Hamilton-Jacobi and Mukhanov-Sasaki equations for scalar perturbations.}

\date{\today}



\author{Thomas Cailleteau, Aurelien Barrau}
\affiliation{Laboratoire de Physique Subatomique et de Cosmologie, UJF, CNRS/IN2P3, INPG\\
53, av. des Martyrs, 38026 Grenoble cedex, France}

\begin{abstract}
Gauge invariance of scalar perturbations is studied together with the associated equations of motion. Extending methods developed in the framework of hamiltonian general 
relativity, the Hamilton-Jacobi equations are investigated into the details in Loop Quantum Cosmology. The gauge-invariant observables are built and their equations of
motions are reviewed both in Hamiltonian and Lagrangian approaches. This method is applied to scalar perturbations with either holonomy or inverse-volume corrections. 
\end{abstract}

\maketitle


\section{Introduction}

Loop Quantum Gravity (LQG) is a non-perturbative and background-independent 
quantization of General Relativity (GR) (see \cite{lqg_review} for reviews). Recently, it has been realized that different
views, based on canonical quantization of GR, on covariant quantization of GR and on formal quantization 
of geometry lead to the very same LQG framework. Although other approaches are still much debated, this makes LQG a very promising model
to address the outstanding question of quantum gravity. 

The application of LQG ideas to the universe as a whole is called Loop Quantum Cosmology (LQC) (see \cite{lqc_review} for reviews).
This is basically the symmetry-reduced version of the theory. So far, LQC proved to be interesting both as a model of the early universe, solving
the Big Bang singularity, and as a way of possibly testing LQG ideas. At the effective level, LQC modifies the usual paradigm by two
main corrections: the inverse-volume terms, basically arising for inverse powers of the densitized triad, which when 
quantized become an operator with zero in its  discrete spectrum thus lacking a direct inverse, and holonomy corrections coming
from the fact that loop quantization is based on holonomies, rather than direct connection components.

To investigate the observational consequences of those LQC-induced modifications, it is most useful to construct rigorously gauge-invariant variables.
It is well known, even in standard GR, that among the solutions of field equations for perturbed variables, some are unphysical modes corresponding
to a mere coordinate transformations. \\

In this article, we basically extend the method introduced in \cite{Langlois}. We start with first order constraints in the FLRW metric. Then, using a generating function,
the variables (perturbations of the densitized triads
and their conjugate momenta) are changed according to  $(\delta K, \delta E) \rightarrow (\gamma_m,\pi_m)$. The first order constraints are re-expressed
with respect to $(\gamma_m,\pi_m)$. The gauge-invariant variables $(Q,P)$ are obtained thanks to a natural generating function $S$ and the dynamics is derived
through anomaly-free second order constraints in terms of $(Q,P)$. Then, the Mukhanov
variables $v$, $R$ and $z$ are given. Finally, the method is applied to the case of LQC with
both holonomy and inverse-volume corrections.\\

This approach exhibits several advantages:
\begin{itemize}
\item the treatment is purely Hamiltonian with easy computations,
\item the Mukhanov variables $v$ and $R$ are obtained directly and the equation of motion is easily found without using Bardeen Potentials,
\item it helps to construct an anomaly-free algebra by imposing relations on the Poisson brackets,
\item the $z$ variable can be found without ambiguity and in a quite simple way,
\item the generating functions are clearly defined, easy to handle and allow one to trace back deeply the origin of gauge invariance,
\item it works for any kind of constraint theory. 
\end{itemize}

The paper is organized as follows. In the two first sections, we  introduce the 
framework of Loop Quantum Cosmology and some elements of analytical mechanics
useful to implement the Hamilton-Jacobi method. Then, we show the main steps of the 
proposed procedure and its application to the cases of holonomy and inverse-volume 
corrections.

\section{Loop Quatum Cosmology framework}

In General Relativity, when the ADM formalism is chosen, space-time is foliated into a family of spacelike 3-surfaces and the dynamics is given by constraints. The fundamental variables
are the space metric $q_{ab}$, together with $N$, the lapse function, and $N^a$, the shift vector, which describe how each of the "leaves" of the foliation are welded together. The metric
is written as:
\begin{equation} \lb{generalmetric}
ds^2 = - N^2 dt^2 + q_{ab} (dx^a + N^a dt)(dx^b + N^b dt).
\end{equation}
In the  LQC formalism, the spatial metric is expressed in terms of 
triads $e^i_a$  that are related to the spatial metric by:
\begin{equation}
q_{ab} \dot{=} e^i_a e^i_b.
\end{equation}
The first basic variable (for a detailed introduction, see \cite{lqg_review}) is the Ashtekar
 connection:
\begin{equation}
A^i_a = \Gamma^i_a + \gamma K^i_a,
\end{equation}
where $\gamma $ is the Barbero-Immirzi parameter, $\Gamma^i_a$ is the spin connection and 
$K^i_a$ is the extrinsic curvature. The second one is the densitized triad:
\begin{equation}
E^a_i = ( det \hskip 0.1truecm  e^i_a ) \hskip 0.1truecm e^a_i.
\end{equation}
The conjugate variables follow the symplectic structure
\begin{equation}
\{A^i_a(x), E^b_j(y)\} = \kappa \gamma \delta^{i}_{j} \delta^{b}_{a} \delta^3(x-y),
\end{equation}
where $\kappa = 8 \pi \mathcal{G}$. The canonical Einstein-Hilbert action in this formalism reads as
\begin{equation}
S_{EH} = \int dt \left[\int \frac{d^3x}{\kappa \gamma} \dot{A}^i_a E^a_i - G[\Lambda^i]  -
D[N^a] - H[N] \right],
\end{equation}
where $G[\Lambda^i]$ is the Gauss constraint, $D[N^a]$ is the diffeomorphism constraint, and $H[N]$ is the Hamiltonian constraint. The diffeomorphism constraint generates
deformations of a spatial slice so that, when it is satisfied, spatial geometry does not depend on the choice of space coordinates. General covariance of the spacetime geometry (including the time coordinate) is ensured by the Hamiltonian constraint. Finally, as a set of triad vectors can be rotated without changing the metric, there is an additional SO(3) gauge freedom. Invariance of the 
theory under those rotations is guaranteed by the Gauss constraint.
This latter constraint will be solved explicitly by the parametrization we use for the variables. \\ 

Taking into account perturbations in a FLRW universe, one has to deal with the perturbed 
spatial metric $\delta \gamma_{ab}$ such that:
\begin{equation}
q_{ab} = a^2 (\delta_{ab} + \delta \gamma_{ab}),
\end{equation}
where $a(t)$ is the scale factor. The perturbed Ashtekar variables will then be related to the perturbed metric and
it is straightforward to see that the background and perturbed densitized triad obeying
$E^a_i = \bar{E}^a_i + \delta E^a_i$ 
are:
\begin{equation}
\bar{E}^a_i = \barp \delta ^a_i = a^2 \delta ^a_i,
\end{equation}
and 
\begin{equation} \lb{deE}
\delta E^a_i = \frac{1}{2} \barp \left( - \delta \gamma_i^a + \delta \gamma^d_d \cdot \delta^a_i \right).
\end{equation}
On the other hand, the extrinsic curvature $K^i_a$ is given by
\begin{equation}
K^i_a = \bar{K}^i_a + \delta K^i_a  = \bark \delta ^i_a + \delta K^i_a.
\end{equation}
The homogeneous and isotropic background $(\barp, \bark)$  satisfies 
\begin{equation}
\{\bark, \barp\} = \frac{\kappa}{3},
\end{equation}
and the perturbed part $(\delta E^a_i, \delta K^b_j)$ fulfills
\begin{equation}
\{\delta K^i_a(x), \delta E^b_j(y)\} = \kappa \delta^{i}_{j} \delta^{b}_{a} \delta^3(x-y).
\end{equation}

In \cite{Langlois}, the original variables are the spatial perturbed metric $\delta \gamma_{ab}$ and its conjugate momentum $\delta \pi^{ab}$. 
As $\delta E^a_i$ is linear in $\delta \gamma$, as it can be seen in Eq. \ff{deE}, it is possible to follow the 
same procedure, with only minor modifications due to the fact that now the
fondamental variables are $\delta E$ and $\delta K$. \\ 
The study of the homogeneous and isotropic universe is an important first step for any tentative theory of 
quantum cosmology. In the framework of LQC, this led to the famous replacement
of the Big Bang by a Big Bounce. Investigating perturbations is the next logical step to probe possible 
deviations from the standard model. This has already been studied in many articles 
(see, {\it e.g.}, \cite{tensor}), especially for gravitational waves and subsequent consequences of the B-mode 
spectrum of the Cosmological Microwave Background (CMB). 

We now turn to the study
of scalar perturbations of the metric ($\phi$, $\psi$, $B$ and $E$) that are
 observationally  
relevant 
as they can be used to compute the well measured temperature CMB spectrum. The perturbed FLRW metric in conformal time 
can be written as 
\begin{eqnarray}
ds^2 &=& a^2(\eta) \left[ -(1+2 \phi) d\eta^2+ 2  \partial_a B \cdot dx^a d\eta  \right. \nonumber \\ 
&& \left. ~~~~~ + ((1-2 \psi)\delta_{ab} + 2 \partial_a \partial_b E )  dx^a dx^b \right]. \lb{FLRWmetric}
\end{eqnarray}
Comparing this expression with Eq. \ff{generalmetric}, one obtains the perturbed lapse function and perturbed shift vector as: 
\begin{equation}
\delta N = \barN \phi ; \hskip 1truecm \delta N^a = \partial^a B.
\end{equation}
Using the definition of the densitized triad, one can also see that
\begin{equation} \lb{perturbeddeltaE}
\delta E^a_i = \barp \left(- 2 \psi \delta^a_i + (\delta^a_i \Delta -\partial_i \partial^a )E \right).
\end{equation}
Starting from Eq. \ff{perturbeddeltaE}, all the useful equations will be derived using a clear algorithm. 

\section{Hamilton-Jacobi equation}

This section is heavily based on \cite{Goldstein_meca}.

\subsection{Canonical transformations}
When dealing with general transformations of coordinates, one has to consider the simultaneous transformations of independent
 coordinates and momenta, $q_i$ and $p_i$ to a new set $Q_i$ and $P_i$, through (invertible) equations :
 \begin{eqnarray}
 Q_i &=& Q_i (q,p,t), \\
 P_i &=& P_i(q,p,t).
 \end{eqnarray}
 Theses equations define a transformation from a point in phase space to another one. \\
 In the Hamiltonian mechanics framework, only those transformations  for which the new
 Q,P are canonical coordinates are interesting. This means that there exists a function
 $K(Q,P,t)$ such that the equations of motion in the new set are in the Hamiltonian form:
 \begin{equation} \lb{eqmotion1}
\dot{Q}_i = \frac{\partial K}{\partial P_i}, \hskip 1truecm \dot{P}_i= - \frac{\partial K}{\partial Q_i}.
\end{equation}
Transformations for which Eqs. \ff{eqmotion1} are fulfilled are said to be \textit{canonical}.\\
The function $K$ plays the role of the Hamiltonian in the new coordinate set. For the treatment
 to be
fully generic, for all
systems with the same number of degrees of freedom, Eqs. \ff{eqmotion1} must be the equations of
motion in the new coordinates and momenta whatever the initial form of $H$. No matter wether one 
deals with an
harmonic oscillator or with a two-dimensional Keplerian problem. \\
If $Q_i$ and $P_i$ are to be canonical coordinates, they must satisfy the "modified" 
Hamilton principle
\begin{equation}
\delta \int (P_i \dot{Q}_i - K(Q,P,t))dt = 0,
\end{equation}
whereas, as usual, 
\begin{equation}
\delta \int (p_i \dot{q}_i - H(q,p,t))dt = 0.
\end{equation}
Both equations will be satisfied if the integrands are connected by the relation:
\begin{equation}
\lambda (p_i \dot{q}_i - H(q,p,t)) = P_i \dot{Q}_i - K(Q,P,t) + \frac{dF}{dt},
\end{equation}
and both sets verify the Poisson Bracket:
\begin{equation}
\{q,p\} = \{ Q,P \} = 1.
\end{equation}
$F$ is useful when mixing half of the old variables with the new variables and will then be considered
as a bridge between the two sets of canonical variables. It is called the \textit{generating function} of the
transformation. On can define 4 such generating functions:
\begin{eqnarray}
F_1(q,Q,t), \hskip 1truecm F_2(q,P,t), \nonumber \\
F_3(p,Q,t), \hskip 1truecm F_4(p,P,t),
\end{eqnarray}
with the following properties:
\begin{eqnarray}
&& p_i = \frac{\partial F_1}{\partial q_i}, \hskip 1truecm P_i = - \frac{\partial F_1}{\partial Q_i}, \\
&& K = H +\frac{ \partial F_1}{\partial t},
\end{eqnarray}

\begin{eqnarray}
&& p_i = \frac{\partial F_2}{\partial q_i}, \hskip 1truecm Q_i =  \frac{\partial F_2}{\partial P_i}, \\
&& K = H +\frac{ \partial F_2}{\partial t},
\end{eqnarray}

\begin{eqnarray} \lb{F3eqmvt}
&& q_i =-  \frac{\partial F_3}{\partial p_i}, \hskip 1truecm P_i = - \frac{\partial F_3}{\partial Q_i}, \\
&& K = H +\frac{ \partial F_3}{\partial t},
\end{eqnarray}

\begin{eqnarray}
&& q_i = - \frac{\partial F_4}{\partial p_i}, \hskip 1truecm Q_i =  \frac{\partial F_4}{\partial P_i}, \\
&& K = H +\frac{ \partial F_4}{\partial t}.
\end{eqnarray}
In the particular case where, for instance, 
\begin{equation} \lb{F3Id}
F_3(p,Q,t) = - p \cdot Q,
\end{equation}
Eq. \ff{F3eqmvt} gives 
\begin{equation}
q = Q, \hskip 1truecm P=p,
\end{equation}
which corresponds to the identity transformation.

\subsection{Hamilton-Jacobi equation}

In order to solve a problem of mechanics, it is useful to formulate it with the best suited variables, for example the angle-action variables. Then, one solves the
Hamilton-Jacobi equation written thanks to a generating function, $S$, which changes initial variables to new appropriate coordinates. The equation is basically given by
\begin{equation}
H\left(q_i, p_i = \frac{\partial S}{\partial q_i}  \right)  = \alpha_i.
\end{equation} 
For gravity, one has to deal with constraints, as introduced in the previous 
sections. General methods to solve the Hamilton-Jacobi equation in this case, with $\alpha_i
=0 $, are given in \cite{Goldberg}.

\section{First change of variables $(\gamma_m,\pi_m)$ }
In this section, we show in some details the way to proceed in order to find easily the gauge-invariant quantities. This "algorithm" of resolution, originally used in
\cite{Langlois} but not fully detailed, can be applied in many situations where perturbations are considered (see for instance \cite{gr-qc/0210078}).
The Hamilton-Jacobi framework has already been extensively studied and used in 
General Relativity, as, {\it e.g.}, in \cite{gr-qc/0210078}, but 
the method presented here focuses on the goal of directly deriving
some gauge-invariant variables useful for observations.

 It will here  be explains for the case of General Relativity
but expressed with variables that can be further used in the framework of Loop Quantum Gravity, as
investigated in the last section of this article.  

\subsection{new variables}
Following \cite{Langlois}, we will define, from an appropriate generating function, two "new" variables $\gamma$ and their conjugate momenta $\pi$, 
related to $\delta K$ and $\delta E$, so that the equations get simplified. This is nothing else than re-expressing the perturbations $E$ and
$\psi$. Fundamentally, this does not bring any new physical information as it is possible to obtain the very same results starting directly from $\psi$ and $E$ .\\
We will use the Fourier transformed variables such that, for instance,
\begin{equation} \lb{fourtrans}
\delta E^a_i (k,t) = \int d^3x \hskip 0.1truecm e^{-i {\bf k}\cdot {\bf x} } \hskip 0.1truecm \delta E^a_i(x,t),
\end{equation}
leading to:
\begin{equation}
\delta E^a_i(k,t) = \int d^3x e^{-i k\cdot x} \barp \left(  -2 \psi ( \delta^a_i )  + (\delta^a_i -
\frac{k_i k^a}{k}) k^2 E \right) .
\end{equation}
Working in the Fourier space greatly simplifies most equations and can add some freedom.
In our case, we define two vectors $A^m$ (m=1,2) in the Fourier space such that:
\begin{eqnarray}
A^1_{ai}&=&  a \delta^a_i,  \\
A^2_{ai} &=& b \left( \delta^a_i - \frac{k_i k^a}{k} \right).
\end{eqnarray}
The variables $a$ and $b$ depend on the choice of the basis but, as we will show later, the final results do not depend on them.
The scalar product of these vectors is  proportional to $2 k^2$. This is in sharp contrast with the situation studied in \cite{Langlois} where one had:
\begin{eqnarray}
A^1_{L}&=&  \delta^a_i,  \\
A^2_{L} &=& \frac{k_i k^a}{k} - \frac{1}{3} \delta^a_i.
\end{eqnarray}
The difference is due to the choice of the perturbation: in our case we use $E$, whereas the 
"standard" $\mu$ was used in \cite{Langlois} 
leading to:
\begin{equation}
\delta E^a_i(k,t) = \int d^3x e^{-i k\cdot x} \barp \left(  -2 \psi ( \delta^a_i )  + (\frac{k_i k^a}{k} - \frac{1}{3} \delta^a_i) k^2 \mu \right). \nonumber \\
\end{equation}
It is easy to see that $A^1_L \cdot A^2_L = 0$. However, this is not in principle necessary  and both approaches are strongly related and lead to the same results. 
Having defined theses vectors, instead of working with $\psi$ and $E$, we will use two other variables $\gamma_m$ $(m=1,2)$ such that:
\begin{eqnarray} \lb{deltaE} 
&&\delta E^a_i(k,t) = \gamma_1 A^1_{ai} + \gamma_2 A^2_{ai}.
\end{eqnarray}
As suggested before, these new variables are just related to the perturbations through 
\begin{eqnarray}
a \gamma_1 &=&  2 \barp \psi, \\
b \gamma_2 &=& \barp k^2 E.
\end{eqnarray}
Using Eq. \ff{deltaE}, one can express them in terms of $\delta E$ such that:
\begin{eqnarray}
&&a\gamma_1 =  \frac{k_a k^i}{k^2} \delta E^a_i \lb{gamma1solv}, \\
&&b\gamma_2 = -\frac{1}{2}\left( 3 \frac{k_a k^i}{k^2}\delta E^a_i - \delta E^d_d \right). \lb{gamma2solv}
\end{eqnarray}
Taking the trace of \ff{deltaE} indeed leads to:
\begin{equation}
\delta E^d_d = 3 a \gamma_1 + 2 b \gamma_2,
\end{equation}
and expressing $\gamma_1$ as a function of $\gamma_2$  in Eq. \ff{deltaE}, one obtains:
\begin{equation} 
\delta E^a_i = \frac{1}{3} \delta E^d_d \delta ^a_i + b \gamma_2 \left( \frac{1}{3} \delta^a_i - \frac{k_i k^a}{k^2}  \right).
\end{equation} 
This can be expressed as:
\begin{equation} 
\delta E^a_i = \frac{1}{3} \delta E^d_d A^1_L + b \gamma_2 A^2_L,
\end{equation} 
by replacing $\delta \gamma_{ij}$ (as used in \cite{Langlois}) by $\delta E^a_i$. This is the first bridge between the two approaches. 
When solving this equation  by multiplying  by $A^2_L{}^{-1}$, one naturally obtains Eqs. \ff{gamma1solv} and \ff{gamma2solv}. \\
Furthermore, we can show that both approaches are in fact equivalent. Starting from one, for example using $\mu$,
 we can derive the expression of $\gamma_m$ when $E$ is used in terms of $\gamma_m^L$. This is simply performed by using the relation Eq. \ff{deE} which relates $\delta \gamma_{ij}$
 to $\delta E^a_i$. 
 To be consistent 
with \cite{Langlois}, we re-define (due to our conventions) the variables as:
\begin{eqnarray} \lb{deG}
\barp \delta \gamma^a_i = \barp \gamma_1^L A^1_L + \barp \gamma_2^L A^2_L.
\end{eqnarray}
Noticing that, from Eq. \ff{deE},
\begin{equation} \lb{deG2}
\barp \delta \gamma^a_i = \delta E^d_d \delta ^a_i - 2 \delta E^a_i,
\end{equation}
we can re-express the approach of \cite{Langlois} as:
\begin{eqnarray} \lb{gamma1L}
\gamma_1^L &=& \frac{1}{3} \delta \gamma^d_d, \\
\gamma_2^L &=& \frac{1}{2} \left( 3 \frac{k^i k_a}{k^2} \delta \gamma^a_i - \delta^i_a \delta \gamma^a_i \right).  \lb{gamma2L}
\end{eqnarray}
From Eqs. \ff{deE}, \ff{deltaE}, \ff{deG} and \ff{deG2}, it follows that
\begin{eqnarray} \lb{eq1}
\barp \delta \gamma^a_i &=& \barp \gamma_1^L \delta^i_a +\barp \gamma_2^L  \left( \frac{k_i k^a}{k} - \frac{1}{3} \delta^a_i \right)  \\
&=& a \gamma_1 \delta^a_i + 2 b \gamma_2 \frac{k^i k_a}{k^2}.
\end{eqnarray}
Taking the trace gives:
\begin{equation}
\barp \gamma_1^L = a \gamma_1 + \frac{2b}{3} \gamma_2,
\end{equation}
and  Eq.\ff{eq1} becomes
\begin{equation}
\barp \gamma_2^L  \left( \frac{k_i k^a}{k} - \frac{1}{3} \delta^a_i \right) = 2b \gamma_2  \left( \frac{k_i k^a}{k} - \frac{1}{3} \delta^a_i \right).
\end{equation}
This leads to the expected equations:
\begin{eqnarray}
2 b \gamma_2 &=& \barp \gamma_2^L,  \\
a \gamma_1 &=& \barp \left(\gamma_1^L - \frac{1}{3} \gamma_2^L \right).
\end{eqnarray}
Re-expressing them as functions of $\delta E$, with Eq. \ff{deG2}, \ff{gamma1L} and \ff{gamma2L} leads to Eqs. \ff{gamma1solv} and \ff{gamma2solv}.

\subsection{generating function $S_\gamma$}

In the framework of perturbed LQC, the canonical variables are $\delta K$ and  their conjugates $\delta E$. We have seen that it is possible to make a
transformation from $\delta E$ to $\gamma_m$. To this aim, we have to define the corresponding conjugate variables $\pi_m$ of $\gamma_m$, which will depend on $\delta K$.
As reminded above, there exists 4 generating functions allowing one to define new sets of variables. In our case, we  define the momentum with a 
generating function $S_\gamma$ such that 
\begin{equation}
\pi_m \propto \frac{\partial S_\gamma}{\partial \gamma_m},
\end{equation}
where $S_\gamma$ is then a function of $\gamma_m$. As $\gamma$ is a function of $\delta E$, the momenta do not depend on $\delta E$ and we can therefore set 
\begin{equation}
S_\gamma = c \delta K^i_a A^m_{ai} \gamma_m,
\end{equation}
where $c$ is a constant. At this stage, we might consider two cases. First, one may chose to have $\gamma_m$ as  canonical coordinates and $\pi_m$ as their conjugate momenta:
\begin{eqnarray}
&& \delta K = q, \hskip 1truecm \delta E = p, \nonumber\\
&& \gamma_m = Q, \hskip 1truecm \pi_m = P.
\end{eqnarray}
In this case, $S_\gamma$ will be similar to a $f_1(q,Q)$ function and the conjugate momenta are:
\begin{equation} \lb{firstcase}
\pi_m = - \frac{\partial S_\gamma}{\partial \gamma_m} = - c \delta K^i_a A^m_{ai}.
\end{equation} 
In the second case, as $\gamma_m$ are related to $\delta E = p$, one might want to have now $\pi_m$ as canonical coordinates and $\gamma_m$ as their conjugate momenta:
\begin{eqnarray}
&& \delta K = q, \hskip 1truecm \delta E = p, \nonumber\\
&& \gamma_m = P, \hskip 1truecm \pi_m = Q.
\end{eqnarray}In this case,
$S_\gamma$ will be similar to a $f_2(q,P)$ function and the conjugate momenta are:
\begin{equation} \lb{secondcase}
\pi_m = \frac{\partial S_\gamma}{\partial \gamma_m} =  c \delta K^i_a A^m_{ai}.
\end{equation} 
Comparing both cases, one can see that changing $c \rightarrow -c$ exchanges one case for the other one. From now on, we focus on the first case: the $\gamma_m$ will be the canonical coordinates, and $\pi_m$ their conjugate momenta. Nevertheless,
considering either $\gamma = Q$ or $\gamma = P$ , the algorithm naturally leads to the same correct gauge-invariant
variables. In our choice, those variable are precisely the Mukhanov variables.  \\
Remaining as general as possible, one can finally write:
\begin{eqnarray} \lb{pimomenta}
\pi_1 &=& - a c \delta K^d_d, \\
\pi_2 &=& - c b \left( \delta K^d_d - \frac{k^a k_i}{k^2} \delta K^i_a \right).
\end{eqnarray} 
It is useful for the following computations to re-express $\delta K$ as a function of $\pi_m$ such that
\begin{equation} \lb{deltaK1}
\delta K^i_a = a^i_a \pi_1 + b^i_a \pi_2,
\end{equation} 
where
\begin{eqnarray}
a^i_a &=& a_1 \delta^i_a + a_2 \frac{k^i k_a }{k^2}, \\
b^i_a &=& b_1 \delta^i_a + b_2 \frac{k^i k_a }{k^2}.
\end{eqnarray}
Multiplying Eq. \ff{deltaK1} by $-c A^m_{ia}$ leads to conditions on $a_1$, $b_1$, $a_2$ and $b_2$ through
\begin{equation}
\pi_m = - c A^m_{ia} \left( a^i_a \pi_1 + b^i_a \pi_2 \right),
\end{equation}
and, consequently,
\begin{equation} \lb{deltaK}
\delta K^i_a = - \frac{1}{c a} \frac{k^i k_a}{k^2} \pi_1 + \frac{1}{ 2 b c } \left( 3 \frac{k^i k_a}{k^2} -\delta^i_a \right) \pi_2.
\end{equation}
It is useful to multiply the previous equation by $\delta E^a_i$ so as to obtain:
\begin{eqnarray}
\delta K^i_a \delta E^a_i &=&  - \frac{ \delta E^a_i}{c a} \frac{k^i k_a}{k^2} \pi_1 + \frac{ \delta E^a_i}{ 2 b c } \left( 3 \frac{k^i k_a}{k^2} -\delta^i_a \right) \pi_2
\nonumber \\
&=& -\frac{1}{c} \left( \gamma_1 \pi_1 + \gamma_2 \pi_2 \right) \lb{conservationrelation}.
\end{eqnarray} 
As explained in the next section, this might be interpreted as a conservation equation.\\

\subsection{Poisson brackets}
We now have defined the conjugate momentum $\pi_m$ of $\gamma_m$. As for the original Ashtekar variables, these new ones will obey some Poisson bracket
relations. Going through the computation leads to:
\begin{equation}
\{ \gamma_m, \pi_n \} = \kappa \hk c \hk \delta_{mn}.  
\end{equation}
The transformation can be said to be canonical as the variables have a symplectic structure such that new Poisson bracket is related to the old one through
\begin{equation}
\{ \gamma_m, \pi_n \} = \{\delta K^i_a (x), \delta E^a_i (x) \} \hk c \hk \delta_{mn}.  
\end{equation}
In the next sections, we will consider a universe filled with matter, and, in particular, with a massive scalar field $\bar{\varphi}$, with its conjugate momentum
$\bar{\pi}$, so that $\{ \bar{\varphi}, \bar{\pi} \} = 1 $. For simplicity and without any lack of generality, we therefore set $ \kappa c=  1 $. So, 
\begin{equation}
\{ \gamma_m, \pi_n \} =\delta_{mn}.  
\end{equation}

Using Eq. \ff{conservationrelation}, we see that 
\begin{eqnarray}
\frac{\delta K^i_a \delta E^a_i}{\{\delta K^i_a (x), \delta E^a_i (x) \} }  &=&  -\frac{ \gamma_m \pi_m }{\{ \gamma_m, \pi_m \} }.
\end{eqnarray} 
As we have chosen the $\gamma_m$, related to $\delta E$, as the canonical coordinates (instead of $\delta K$ in the usual theory), this leads to the appearance of a minus sign in the previous equation. \\
The new set of symplectic variables, $\gamma_m$ and $\pi_m$, is now well defined as:
\begin{eqnarray}
&&a\gamma_1 =  \frac{k_a k^i}{k^2} \delta E^a_i \lb{gamma1solv2}, \\
&&b\gamma_2 = -\frac{1}{2}\left( 3 \frac{k_a k^i}{k^2}\delta E^a_i - \delta E^a_i \delta^i_a \right) \lb{gamma2solv2},
\end{eqnarray}

\begin{eqnarray} \lb{pimomenta1}
\pi_1 &=& - ac \delta K^d_d, \\
\pi_2 &=& - bc \left( \delta K^d_d - \frac{k^a k_i}{k^2} \delta K^i_a \right).\lb{pimomenta2}
\end{eqnarray}

Their Poisson brackets are given by:
\begin{eqnarray}
\{\gamma_1 , \pi_1 \} &=& \{ \gamma_2, \pi_2 \} = 1, \\
\{\gamma_1, \pi_2 \} &=& \{ \gamma_2 , \pi_1 \} = 0.
\end{eqnarray}
As explained before, at this stage, nothing new emerges. This transformation is just useful to obtain simpler equations. 

\subsection{First order constraints in term of $(\gamma_a, \pi_a)$}
Gauge-invariant variables are derived from first order constraints. To use our new set of variables, we now have to re-express the
constraints in terms of $\gamma_m$ and $\pi_m$. With the Ashtekar variables, for a universe filled with a massive scalar field $\varphi$ 
(with conjugated momentum $\pi$), the diffeomorphism constraints in the ADM formalism reads as
\begin{equation} \lb{diffeocons}
D[N^a] = \int_\Sigma d^3x \left[ \barN^a (\mathcal{D}^{(0)} + \mathcal{D}^{(2)}) + \delta N^a \mathcal{D}^{(1)} \right],
\end{equation}
at first order in constraint densities. In fact, as in this case $\barN^a = 0$, only the $\mathcal{D}^{(1)}$ term remains. Its gravitational and matter
components are:
\begin{equation} \lb{densDinit}
\mathcal{D}^{(1)}_{G} = \frac{1}{\kappa} \left(  -\bark \delta^k_c \partial_d(\delta E^d_k) +  \barp  \partial_c(\delta K^d_d)- \barp  \partial_d(\delta K^d_c) \right),
\end{equation}
\begin{eqnarray}
\mathcal{D}^{(1)}_{M} =  \bar{\pi} (\partial_c \delta \varphi).
\end{eqnarray}

As far as the Hamiltonian constraints are concerned, one has: 
\begin{equation}
H[N] = \int_\Sigma d^3x \left[ \barN (\mathcal{H}^{(0)} + \mathcal{H}^{(2)}) + \delta N \mathcal{H}^{(1)} \right],
\end{equation}
where the first order constraint densities are:
\begin{equation} 
\mathcal{H}^{(1)}_{G} = \frac{1}{2 \kappa} \left(  -4 \sqrt{\barp} \bark  \delta K^d_d - \frac{1}{\sqrt{\barp}} \bark ^2 \delta E^d_d + 
\frac{2}{\sqrt{\barp}}  \partial_c \partial^j \delta E^c_j \right),
 \end{equation}
\begin{eqnarray}
\mathcal{H}^{(1)}_{\pi} &=&  \frac{\bar{\pi} \delta \pi}{\bar{p}^{3/2}}-\frac{\bar{\pi}^2}{2\bar{p}^{3/2}} \frac{\delta^j_c \delta E^c_j}{2\bar{p}},   \\
\mathcal{H}^{(1)}_{\varphi} &=& 
\bar{p}^{3/2} \left[ V_{,\varphi}(\bar{\varphi}) \delta \varphi +V(\bar{\varphi})
\frac{\delta^j_c \delta E^c_j}{2\bar{p}} \right]. \lb{densHfin}
\end{eqnarray}
As in \cite{Langlois}, we define $\gamma_0 = \delta \varphi$, and $\pi_0 = \delta \pi$, such that $\gamma_a$ ($a=0,1,2$) correspond to the old canonical coordinates "q" and
$\{\gamma_a, \pi_a\} = 1$. The expressions of $\gamma_a$ represent the "maximal" and 
"fundamental" decomposition of the perturbations. What was
done so far is nothing else than a decomposition of the theory in terms of those perturbations. \\
After a Fourier transformation, and using Eqs. \ff{deltaE} and \ff{deltaK}, both first order constraints  can now be expressed as functions of $(\gamma_a, \pi_a)$ such that:
\begin{eqnarray} \lb{H1E}
&&\mathcal{H}^{(2)}[\delta N] = \delta N  \left( \mathcal{H}^{(1)}_{G} + \mathcal{H}^{(1)}_{M} \right) \nonumber \\
&& =   \frac{\delta N}{\sqrt{\barp}} \left[ \frac{2 \barp \bark}{c \kappa } \frac{\pi_1}{a} + \dot{\bar{\varphi}}  \pi_0 +
 \frac{b \gamma_2 }{\kappa} \left(2 \bark^2 - \kappa \dot{\bar{\varphi}}^2 \right) \right. \nonumber \\
&& \left. +  \barp^2 V' \gamma_0 + \frac{a \gamma_1}{\kappa} \left(- k^2 + 3\bark^2- \frac{3}{2}\kappa \dot{\bar{\varphi}}^2\right)  \right],
\end{eqnarray}
and 
\begin{eqnarray} \lb{D1M}
&&\mathcal{D}^{(2)}[\delta N^a] = \delta N^a \left( \mathcal{D}^{(1)}_{G} + \mathcal{D}^{(1)}_{M} \right) \nonumber \\
&&= i \barp  (k_a \delta N^a) \left[- \dot{\bar{\varphi}} \gamma_0 + \frac{\bark}{\barp} \frac{a \gamma_1}{\kappa} +  \frac{1}{c \kappa } \frac{\pi_2}{b}  \right],
\end{eqnarray}
where $V'$ and $V"$ refer respectively to the first and second derivative with
respect to the scalar field $\varphi$. The
 notation $\mathcal{H}^{(2)}[\delta N]$ and $\mathcal{D}^{(2)}[\delta N^a]$ are used, in agreement with most papers, to emphasize that those expressions 
are in fact second order ones due to $\delta N$
and $\delta N^a$ factors.  We have also simplified the results by using the Friedmann equation (calculations are derived in the Appendix):
\begin{equation} \lb{friedmannclassic}
\bark^2 = \frac{\kappa}{3} \left( \frac{\dot{\bar{\varphi}}^2}{2} + \barp V \right),
\end{equation} 
and the fact that the equation of motion for the background variables reads as:
\begin{equation}
\dot{\bar{\varphi}} = \frac{\barpi}{\barp}.
\end{equation}
We have thus expressed the first order constraint densities \ff{H1E} and \ff{D1M}, as functions of the new set of symplectic coordinates $(\gamma_a, \pi_a)$. In the next section, we
will show that it is possible to make a last transformation toward a new set of coordinates $(Q,P)$, meaningful for cosmology and gauge-invariant, using a generating function $S$ and the Hamilton-Jacobi equations.

\section{gauge transformation and the Mukhanov-sasaki equation}
To describe physical effects, one has to deal with gauge-invariant quantities. The goal of this section is to address the specific issue of gauge invariance
within the canonical formalism.

\subsection{gauge-invariance in the canonical formalism}
In a canonical formulation, gauge transformations are directly generated by Poisson brackets of the fields with the constraints.
In the covariant language, gauge transformations are studied as perturbation transformations under the coordinate change
\begin{equation}
x^\mu \rightarrow x^\mu{}' = x^\mu + \xi^\mu(x),
\end{equation}
generated by vector fields $\xi^\mu$. Under this coordinate transformation, any tensor field receives a correction given by its Lie derivative along
$\xi^\mu$. As defined in \cite{Bojowald:2008jv}, the part of the transformation relevant for the scalar modes can be parametrized by two scalar functions $\xi^0$ and $\xi$ 
such that
\begin{equation}
\xi^\mu = (\xi^0, \partial^\mu \xi).
\end{equation}
Along this vector, a variable $X$ will undergo a transformation given in the canonical formalism by:
\begin{equation} \lb{defgauge}
\delta_{[\xi^0,\xi]}X \dot{=} \{ X, H^{(2)}[\barn \xi^0] + D^{(2)}[\partial^a \xi]\},
\end{equation}
where 
\begin{eqnarray}
 H^{(2)}[\delta N] &=&  \int d^3x \mathcal{H}^{(2)}[\delta N], \\
 D^{(2)}[\delta N^a]&= &\int d^3x \mathcal{D}^{(2)}[\delta N^a].
\end{eqnarray}
It is easy to relate the canonical approach to the Lie derivative by noticing that:
\begin{equation}
\{\bar{X} + \delta X, D[\xi^a] \} = \mathcal{L}_{\vec{\xi}}\hskip 0.1truecm(\bar{X} + \delta X).
\end{equation}
In the framework of LQC, using the densitized constraints (\ref{densDinit}-\ref{densHfin}) 
in Eq. \ff{defgauge}, one expresses the transformations of basics variables 
as:
\begin{eqnarray} \lb{pertgen}
&&\delta_{[\xi^0,\xi]}\delta E^a_i  =  2\barp \bark \xi_0 \delta^a_i - \barp (\delta^a_i k^2 - k_a k^i)\xi, \\
&&\delta_{[\xi^0,\xi]}\delta K^a_i  =  -\frac{1}{2} \bark^2 \xi_0 \delta ^i_a -k_a k^i(\xi_0 + \bark \xi) \nonumber \\
&& ~~~~~~~~~~~~~~~ + \frac{\kappa}{2} \left(-\frac{\dot{\bar{\varphi}}}{2}  + \barp V \right)\xi_0 \delta^i_a,   \\
&&\delta_{[\xi^0,\xi]}\delta \varphi  = \dot{\bar{\varphi}} \xi_0,\\
&&\delta_{[\xi^0,\xi]}\delta \pi = - \barp \dot{\bar{\varphi}} k^2 \xi - \barp^2 V' \xi_0.
\end{eqnarray}

With these expressions and the definition of $\gamma_m$ and $\pi_m$, it is easy to see that:

\begin{eqnarray*}
\delta_{[\xi^0,\xi]} H^{(2)}[\delta N]   &=& \{ H^{(2)}[\delta N] , H^{(2)}[\barn \xi^0] + D^{(2)}[\partial^a \xi] \} \\ &=&  0, \\
\delta_{[\xi^0,\xi]} D^{(2)}[\delta N] &=& \{ D^{(2)}[\delta N^a] , H^{(2)}[\barn \xi^0] + D^{(2)}[\partial^a \xi] \} \\ &=& 0,   
\end{eqnarray*}

which means that the first order constraints \ff{H1E} and \ff{D1M} are gauge-invariant. Another way to see this is to compute directly the Poisson brackets:
\begin{eqnarray}
\{ H^{(2)}[\delta N_1] , H^{(2)}[\delta N_2] \} &=&0, \lb{H1H1} \\
\{ H^{(2)}[\delta N] , D^{(2)}[\delta N^a] \} &=& 0, \lb{H1D1} \\
\{ D^{(2)}[\delta N^a_1] , D^{(2)}[\delta N^a_2] \} &=& 0, \lb{D1D1}
\end{eqnarray}
and replace the $\delta N$ and $\delta N^a$ by their $\xi^\mu$ equivalents.\\
What is shown here has been noticed in \cite{Langlois}. As it will be emphasized in the next section, this means that to obtain gauge-invariant quantities, 
the algebra should not only be anomaly-free, but should also have a null first
order Poisson bracket.

\subsection{gauge-invariance with the Hamilton-Jacobi equation}

In the Hamilton-Jacobi equation, the momentum is expressed in terms of a generating function $S$ and a new transformation is performed. As stated in 
\cite{Langlois}, there are differences between the classical case where standard Hamiltonians are used, and the case studied here
where we rely on constraints and reduce the phase space. In the latter case, the Hamilton-Jacobi-like equation has to be directly 
solved. 
As the total first order  constraint (density) has to be null for all $\delta N$ and $\delta N^a$, which play similar (although slightly different) roles, one can separate the equations and
 solve the two Hamilton-Jacobi-like expressions: 

\begin{eqnarray} \lb{Enul}
\mathcal{H}^{(2)}[\delta N] \left(\gamma_\alpha, \pi_\alpha = \frac{\partial S}{\partial \gamma_\alpha} \right) &=& 0, \\
\mathcal{D}^{(2)}[\delta N^a] \left(\gamma_\alpha, \pi_\alpha = \frac{\partial S}{\partial \gamma_\alpha} \right) &=& 0. \lb{Mnul}
\end{eqnarray}
Because $\mathcal{H}^{(2)}[\delta N]$ and $\mathcal{D}^{(2)}[\delta N^a]$ are linear in $(\gamma_a, \pi_a)$ in \ff{H1E} and \ff{D1M}, the more "natural"
generating function to consider is a quadratic function $S = f_2(\gamma_a=q,P)$ such that:
\begin{eqnarray} \lb{SF2}
S&=&\frac{1}{2} A_{\alpha \beta} \gamma_\alpha \gamma_\beta + B_\alpha \gamma_\alpha P_2,
\end{eqnarray}
where $A_{\alpha \beta}$ is a $3 \times 3$ matrix. Taking into account the properties of the generating function, the conjugate variable of 
$P_2$ is given by:
\begin{equation} \lb{eqQ}
Q_2 = \frac{\partial S}{\partial P_2} = B_a \gamma_a.
\end{equation} 
To show where the gauge-invariance of the new variables $Q_2$ and $P_2$ comes from, one can synthetically write that
\begin{eqnarray}
&& \mathcal{H}^{(2)}[\delta N] \nonumber \\
&& =(\delta N )\left( E_a \pi_a + \Sigma_b \gamma_b \right) \nonumber \\
&& = (\delta N) \left(( E_a A_{ab} +  \Sigma_b)  \gamma_b + E_a B_a P \right), ~~~ \lb{H2E}
\end{eqnarray} 
and
\begin{eqnarray}
&& \mathcal{D}^{(2)}[\delta N^a] \nonumber \\
&& = i\cdot (k_c \delta N^c ) \left( M_a \pi_a + \Lambda_b \gamma_b \right) \nonumber \\
&& = i\cdot (k_c \delta N^c ) \left(( M_a A_{ab} +  \Lambda_b)  \gamma_b + M_a B_a P \right). ~~~ \lb{D2M}
\end{eqnarray} 
As they are constraints, they have to vanish and considering $P$ and $\gamma_a$ as independent, it is possible to find $B_a$ and $A_{ab}$ through 4 equations:
\begin{eqnarray} \lb{condS1}
&& E_a B_a  = 0, \hskip 1truecm   M_a B_a = 0, \\
&&E_a A_{ab} +  \Sigma_b = 0, \lb{condS2}\\
 &&M_a A_{ab} +  \Lambda_b = 0. \lb{condS3}
\end{eqnarray} 
With the expression of $Q$ given in \ff{eqQ}, and using Eq. \ff{condS1}, one can see that, with our choice of generating function:
\begin{eqnarray}
\delta_{[\xi^0,\xi]} Q_2 &=& B_a \delta_{[\xi^0,\xi]} \gamma_a \nonumber  \\
&=& B_a \{ \gamma_a , \mathcal{H}^{(2)}(\barN \xi_0) +\mathcal{D}^{(2)}(\partial^a \xi) \} \nonumber \\
&=& B_a \{ \gamma_a ,(\barN \xi_0) ( E_a \pi_a + \Sigma_b \gamma_b) \} \nonumber \\
&&  ~~~+ B_a \{ \gamma_a ,(\partial^a \xi)(M_a \pi_a + \Lambda_b \gamma_b) \},\nonumber \\
&=& (\barN \xi_0)\cdot B_a E_a + (\partial^a \xi) \cdot B_a M_a \nonumber \\
&=& 0.
\end{eqnarray}
This shows that $Q_2$ is basically gauge-invariant because of the relations \ff{condS1} and 
not because the anomaly-freedom of the algebra.
The gauge-invariance can also be seen by expressing $Q_2$ in terms of 
$(\delta E, \delta \varphi)$,  and using \ff{pertgen}. 
Finally, it is possible to define a set of 3 new variables $(Q_b, P_b)$, and the function $S$ by $S = \frac{1}{2} A_{ab} \gamma_a \gamma_b + B_{ab}\gamma_a P_b$. 
Making this choice and 
applying the procedure described above leads to simple equations showing that $Q_b \propto Q$. This shows that there is a unique consistent choice for $Q_2$ and the previous case is therefore preferred.  \\
As far as the generating function is concerned,  we could also have chosen 
$S = f_1 (\gamma_a =q, Q)$ and found the conjugate momentum $P$. This would however have led
 to 
$P_1 = - Q_2$ and the situation would have been  equivalent. Moreover, it is also possible to consider a generating function such that 
\begin{equation} 
S= f_{3,4}(p,\{Q,P\}) = \frac{1}{2} A_{ab} \pi_a \pi_b + B_a \pi_a P_{3,4}.
\end{equation}
Following the same procedure would lead to some new gauge-invariant functions. However they do not exhibit any interesting physical feature. In the following, we will therefore focus on the generating function given in \ff{SF2}, leading to $Q = Q_2$ as a Mukhanov variable and $P=P_2$ as its momentum.

\subsection{Anomaly-freedom in the Hamilton-Jacobi approach }
To be consistent, that is to ensure that the evolution generated by the constraints remains compatible with the constraints themselves, the theory must be anomaly-free. 
In our case, using the same synthetic formulation as in \ff{H2E} and \ff{D2M}, one can compute the Poisson brackets:
\begin{eqnarray*}
&&\{ H^{(2)}[\delta N_1] , H^{(2)}[\delta N_2] \} =0,  \\
&&\{ H^{(2)}[\delta N] , D^{(2)}[\delta N^a] \}=(\delta N)(\delta N^a)(\Sigma_a M_a - E_a \Lambda_a), \\
&&\{ D^{(2)}[\delta N^a_1] , D^{(2)}[\delta N^a_2] \} = 0.
\end{eqnarray*}
The total first order constraint $M^{(1)} [\delta N, \delta N^a]$ leads to:
\begin{equation}
\{ M^{(1)}[1], M^{(1)}[2] \}=[\delta N,\delta N^a](\Sigma_a M_a - E_a \Lambda_a).\\
\end{equation}
To close algebra, that is to cancel anomalies, one has to require that
\begin{equation} \lb{condAnomFree}
\Sigma_a M_a - E_a \Lambda_a = 0.
\end{equation}
Using Eqs. \ff{condS2} and \ff{condS3}:
\begin{eqnarray}
\ff{condS2} \times M_b = E_a A_{ab} M_b +  \Sigma_a M_a = 0, \\ 
\ff{condS3} \times E_b = M_a A_{ab} E_b +  \Lambda_a E_a = 0.
\end{eqnarray}
Combining those equations with \ff{condAnomFree}, the condition for anomaly-freedom reads as:
\begin{equation}
 E_a A_{ab} M_b = M_a A_{ab} E_b,
\end{equation}
which is fulfilled only if $A_{ab}$ is a symmetric matrix, with thus only 6 unknown parameters. This corresponds to a fully solvable problem.
The condition of anomaly freedom allows one to completely determine without ambiguity the equations of motion
 for the gauge-invariant variables.

\subsection{Mukhanov equation in General Relativity}

As perturbations can, {\it a priori}, be analyzed through different choices of gauges (for instance, the \textit{Newton Gauge} where $B=E=0$), it is useful to provide 
gauge-invariant quantities (related to $\gamma_a$, $\delta N$ and $\delta N^a$)
that are physically relevant to investigate observational consequences, the \textit{Bardeen Potentials} \cite{Bardeen}:
\begin{eqnarray}
{\bf \Phi } &=&  \phi + \frac{d}{d\eta }  (B - \dot{E})+ \mathcal{H} (B - \dot{E}),  \\ 
{\bf \Psi } &=&  \psi  - \mathcal{H} (B - \dot{E}), 
\end{eqnarray}
where $\mathcal{H}$ is the \textit{conformal Hubble parameter}. As we are dealing with a universe filled with 
a massive scalar field $\varphi$, it will also undergo gauge-invariant perturbations $\delta \varphi^{GI}$. Gravity and matter perturbations are of course linked
and we shall focus on the  linear order, as often in perturbation theory.
As derived in \cite{Bojowald:2008jv} (this follows from the definition \ff{defgauge}):
\begin{equation}
\delta \varphi^{GI} = \delta \varphi + \dot{\bar{\varphi}} ( B- \dot{E}).
\end{equation}
When dealing with all the scalar perturbations, there are 2 degrees of freedoms. This equation generates a constraint and only 1 
degree of freedom remains. Usually, the relevant variable used in cosmology is called the \textit{Mukhanov-Sasaki variable} $v$, originally found in \cite{Mukhanov:1990me} governed by the associated Mukhanov-Sasaki equation:
\begin{eqnarray}
\frac{d^2v}{d\eta^2}+ \Delta v - \frac{\ddot{z}}{z} v &=& 0,
\end{eqnarray}
where
\begin{eqnarray}
v &=& a(\eta) \left[\delta \varphi^{GI} + \frac{\dot{\bar{\varphi}}}{\mathcal{H}} {\bf \Phi }  \right] \lb{mukv},\\
z &=& a(\eta) \frac{\dot{\bar{\varphi}}}{\mathcal{H}}  \lb{mukz}.
\end{eqnarray}
When this variable is found, and after performing a Fourier transform, it is easy to compute the power spectrum of the
 \textit{conserved curvature perturbation}. As review, \textit{e.g.}, in \cite{Jmartin}:
 \begin{equation}
 v = z \cdot R,
\end{equation}
\begin{equation}
\mathcal{P}_R(k) = \frac{k^3}{\pi^2}\left| \frac{v_k}{z} \right|^2.
\end{equation}
The spectral index is, for example, given by:
\begin{equation}
n_s -1 = \left. \frac{d\hk ln(P_R) }{d \hk ln(k) } \right|_{k = k_\star}.
\end{equation}
This power spectrum $P_R(k,\eta)$, typically representing the state of the universe at the end of inflation, is a mandatory input to compute observables, in particular in the Cosmic Microwave Background (CMB).

\section{Second change of variables $(Q,P)$}
In order to compute physical effects, one needs gauge-invariant variables. We have
shown that the generating function, defined in Eq. \ff{SF2}, will give such gauge-invariant observables. In the following, we will precisely show that $Q$ and 
$P$ are the
Mukhanov variables fulfilling the correct equations of motion. 

\subsection{Expression of the gauge-invariant variables}

Using the requirement \ff{Enul} with $\kappa c = 1$ leads to conditions on $A_{ab}$ and $B_{a}$, through \ff{condS2}. They can be written as:
\begin{equation} 
\mathcal{H}^{(2)}[\delta N] = 0 = 1 \cdot \xi + \gamma_0 \cdot \xi_0 + \gamma_1 \cdot \xi_1 + \gamma_2 \cdot \xi_2,
\end{equation}
where 

\begin{eqnarray}
\xi_{} &=& \frac{2 \barp \bark}{a  } B_1 +\dot{\bar{\varphi}} B_0 \lb{cond1}, \\ 
\xi_0 &=& A_{00} \dot{\bar{\varphi}} + \frac{2 \barp \bark }{a } A_{01} + \barp^2 V' \lb{cond2}, \\
\xi_1 &=& A_{01} \dot{\bar{\varphi}} + \frac{2 \barp \bark }{a } A_{11} + \frac{a}{\kappa}(-k^2 + 3 \bark^2 - \frac{3}{2} \kappa \dot{\bar{\varphi}}) \lb{cond3}, ~~~\\
\xi_2 &=& A_{02} \dot{\bar{\varphi}} + \frac{2 \barp \bark}{a } A_{21} + \frac{b}{\kappa} \left(2\bark^2 - \kappa \dot{\bar{\varphi}} \right). \lb{cond4}  
\end{eqnarray}

Considering now Eq. \ff{Mnul}, we are led to:

\begin{equation} 
\mathcal{D}^{(2)}[\delta N^a] = 0 = 1 \cdot \Xi + \gamma_0 \cdot \Xi_0 + \gamma_1 \cdot \Xi_1
+ \gamma_2 \cdot \Xi_2,
\end{equation}
where,
\begin{eqnarray}
\Xi &=& B_2 \lb{cond5}, \\
\Xi_0 &=& b \dot{\bar{\varphi}} -  A_{02} \lb{cond6}, \\
\Xi_1 &=& \frac{ab}{\kappa}\frac{\bark}{\barp} + A_{12} \lb{cond7},\\
\Xi_2 &=& A_{22} \lb{cond8}.
\end{eqnarray}

This system is much simpler than in the ADM formalism and can be explicitly solved.
Taking into account Eq. \ff{Mnul}, one we can directly fix:
\begin{equation} \lb{condM}
B_2 = A_{22} = 0, \hskip 1truecm A_{02} = b \dot{\bar{\varphi}}, \hskip 0.5truecm A_{12} = -\frac{ab}{\kappa}\frac{\bark}{\barp}.
\end{equation} 
This choice for $A_{02}$ and $A_{12}$ leads to consider, in Eq. \ff{Enul}, only 3 equations 
for 5 unknown variables. One can check that the conditions 
\ff{condM}, implemented in Eq. \ff{cond4}, make it vanish. This choice is therefore obviously 
correct. Moreover, as it will be made clear in the following, it
is not necessary 
to determine all the coefficients of $A_{ab}$. Let us now focus on $B_a$. Eq. \ff{cond1} gives one relation between the terms of $B_a$:
\begin{equation} \lb{relationB0B1}
 B_1 = - \frac{\dot{\bar{\varphi}} a }{2 \barp \bark } \cdot B_0.
\end{equation}
$B_0$ will be kept as an irreducible degree of freedom and we will show that any gauge-invariant quantity will just be, at the end, proportional to the fundament $B_0$  choice . 

The new variable $Q$ defined in Eq. \ff{eqQ} can be expressed in terms of $\gamma_a$ such that, finally, 
using $a \gamma_1 =- 2 \barp \psi$, one obtains:
\begin{equation}
Q = B_0 \left( \delta \varphi - \frac{\dot{\bar{\varphi}}}{2 \barp \bark } \cdot a \gamma_1 \right)  = B_0 \left( \delta \varphi + \frac{\dot{\varphi}}{\bark} \psi \right),
\end{equation}
which is similar to $v$ in \ff{mukv} if one chooses $B_0 = \sqrt{\barp} = a(\eta)$. It can also be noticed that it is independent of the choice of the base. 
Of course, one can also choose to invert Eq. \ff{relationB0B1},
\begin{equation}
 B_0 = - \frac{2 \barp \bark}{ a \dot{\bar{\varphi}} } \cdot B_1,
\end{equation}
and define the gauge-invariant variable as
\begin{equation}
Q = B_1 \left( -\frac{2 \barp}{a}  \frac{\bark}{\dot{\bar{\varphi}}} \gamma_0+  \gamma_1  \right)  = 
-\frac{2 \barp B_1}{a} \left( \frac{\bark}{\dot{\bar{\varphi}}}\delta \varphi + \psi  \right),
\end{equation}
which is a function of the perturbed curvature variable R:
\begin{equation}
Q = -\frac{2 \barp B_1}{a} R,
\end{equation}
which depends on $a$. In the following, we will focus on the case  for which $Q \propto v$ 
but the other choice would also be possible and the same method would lead to similar results.
 Moreover, taking $B_0 = 1$ instead of $\sqrt{\barp}$ makes the calculation easier in order to find the function 
$z$, as in \ff{mukz}.  In the next step, we will keep $B_0$ free, until the last step, and derive the equations of motion for $Q$, and therefore for $v$,  its Hamiltonian
formulation and how to find the expression for $z$. 

\subsection{Hamiltonian expression and Equations of Motions}

A general expression has been found for a gauge-invariant quantity $Q$ which is related to the Mukhanov variables. 
As the generating functions
 $S_\gamma$ and $S$ are known, it is possible to find the Hamiltonian, and therefore the Lagrangian,  from which the equation of motion for $Q$ can be derived. The classical results can then be
 obtained elegantly in the canonical formalism. Considering $\gamma_1$ and $\gamma_2$ as pure 
 gauge variables, as explained into the details in \cite{Goldberg}, we should avoid to use 
 any function explicitly depending on 
 them as they do not contribute to the dynamics. 

 As we know the Hamiltonian as a function of $(\delta K, \delta E)$ and as we have derived the expression of the generating 
 functions, it is possible to express the second order constraints, that are governing the 
 dynamics of perturbations, in terms of the new 
  set of variables $(Q,P)$.
Using the notations of \cite{Langlois}, the known variables can be inverted and one can easily find, with $B_0 = f(\bark, \barp)$:
\begin{eqnarray} \lb{inveq}
&& \delta \varphi = \frac{Q}{B_0} + [\gamma_1, \gamma_2], \\
&& \delta \pi = B_0 P + \frac{A_{00}}{B_0} Q + [\gamma_1, \gamma_2],\\
&&  \pi_1 = \frac{A_{01}}{B_0} Q - \frac{a \dot{\bar{\varphi}}}{2 \barp \bark} B_0 P  + [\gamma_1, \gamma_2],\\
&& \pi_2 = b \frac{\dot{\varphi}}{B_0} Q + [\gamma_1, \gamma_2].
\end{eqnarray}

To go further in studying the dynamics, let us notice that, as $Q$ is gauge-invariant,
\begin{equation}
\delta_{\delta N, \delta N^a} Q \dot{=}\{ Q, H^{(2)}[\delta N] + D^{(2)}[\delta N^a]\} = 0.
\end{equation}
The evolution of $Q$ (and this is true for any gauge-invariant variable) is thus given by: 
\begin{eqnarray}
\dot{Q} &=& \{Q, \barN (H^{(0)} + H^{(2)}) + \barN^a (D^{(0)} + D^{(2)}) \nonumber \\
&&  + H^{(2)}[\delta N] + D^{(2)}[\delta N^a] \} \nonumber \\
&= &  \{Q, \barN H^{(0)} + \barN^a D^{(0)} \} + \{Q,  H^{(2)}[\delta N] + D^{(2)}[\delta N^a] \} \nonumber \\
&& + \{Q, \barN H^{(2)} + \barN^a D^{(2)} \} \nonumber \\
&=& 0 + \delta_{\delta N, \delta N^a} Q + \{Q, \barN H^{(2)} + \barN^a D^{(2)} \}.
\end{eqnarray}
So,
\begin{equation}
\dot{Q} = \{Q, \barN H^{(2)} + \barN^a D^{(2)} \}.
\end{equation}

In our case, these constraints are, with $\barN^a = 0$:
\begin{eqnarray}
\mathcal{H}^{(2)} &=& \frac{1}{2 \kappa} \sqrt{\barp} ( \delta^c_k \delta^d_j \delta K^j_c \delta K^k_d  - (\delta K^d_d )^2 ) \nonumber \\
&&+  \frac{1}{2} \frac{\delta \pi^2}{\bar{p}^{3/2}} + \frac{1}{2} \sqrt{\bar{p}} \delta^{ab} \partial_a \delta \varphi  \partial_b \delta \varphi \nonumber \\
&& + \frac{1}{2}  \bar{p}^{3/2} V_{,\varphi\varphi}(\bar{\varphi}) \delta \varphi^2   + [\delta E^a_i].
\end{eqnarray}
As $\delta E$ is related to $\gamma_m$ through Eq. \ff{deltaE}, we should not consider 
 functions depending on it. Moreover, as 
\begin{equation}
\frac{\partial S_\gamma}{\partial \eta} = f(\gamma_m, \dot{\gamma}_m),
\end{equation}
one can write, after taking the Fourier transformation and using Eq. \ff{deltaK}:
\begin{eqnarray} \lb{Hamgampi}
H^S = \int d^3 k \frac{\barp}{2 \kappa} &&\left[ \frac{3}{2} \frac{\pi^2_2}{(b c) ^2} - \frac{2}{c^2}  \frac{\pi_1}{a} \frac{\pi_2}{b} \right. \nonumber \\
&& \left. + \kappa \frac{\pi_0^2}{\barp^2} + \kappa (k^2 + \barp
V")\gamma_0^2 \right].
\end{eqnarray}

The $c$ parameter enters the equation only quadratically, therefore the choice of the generating function $S_\gamma$ (either $\gamma=P$ or $\gamma=Q$) does not enter the final result.
However, we do not have yet expressed the Hamiltonian for the gauge-invariant variables. It is necessary to use Eqs. \ff{inveq} in \ff{Hamgampi} and to add the relevant terms associated with the derivative with respect to time of the generating function, 
\begin{eqnarray}
\frac{\partial S}{\partial \eta} &=& \frac{1}{2} \dot{A}_{00} \gamma^2_0 + \dot{B_0} \gamma_0 P + f(\gamma_m, \dot{\gamma}_m) \nonumber \\
&=& \frac{1}{2} \dot{A}_{00}  \left( \frac{Q}{B_0} \right)^2 +  \frac{\dot{B_0}}{B_0} PQ.
\end{eqnarray}
To avoid inconsistencies, $\dot {Q}$ and $\dot{P}$ terms should not be taken into account. The gauge-invariant hamiltonian constraint for $Q$ and $P$ is thus:
\begin{eqnarray} \lb{HGI1}
H^S_{GI} = && \int \frac{d^3k}{2} \left[ \left(B_0 \frac{P}{\sqrt{\barp}}\right)^2  + \Gamma \left(\frac{\sqrt{\barp} Q}{B_0}\right)^2  \right. \nonumber \\
&& \left.  + PQ \left(\frac{1}{\kappa c^2 }\frac{\dot{\bar{\varphi}}^2}{\bark} + 2 \frac{A_{00}}{\barp} + 2 \frac{\dot{B_0}}{B_0}\right)\right],
\end{eqnarray}
where 
\begin{eqnarray} \lb{Gamma1}
 \Gamma = \frac{\dot{A}_{00}}{\barp} &-& 2 \kappa \dot{\bar{\varphi}} \frac{A_{01}}{a} +\frac{A^2_{00}}{\barp^2} \nonumber \\
 && +  k^2 + \barp V" + \frac{3 \kappa }{2} \dot{\bar{\varphi}}^2.
\end{eqnarray}
In order to recover the usual Hamiltonian formulation, the cross terms in $P$ and $Q$ in the previous equation should vanish. However, looking at \ff{HGI1}, it is clear that this cross terms will give the expression of $A_{00}$, as a function of $\frac{\dot{B_0}}{B_0}$. However, looking at $\Gamma$, we see that the square and the derivative with respect to  time 
of $A_{00}$ are involved. For $B_0 \neq 1$, the expression is quite complicated (although in
principle tractable). As we are interested
in the Mukhanov variable $v$, an easy choice is to
first consider $B_0 = 1$, to find the equation of motion for $Q$ and its Hamiltonian, and then to set $v = \sqrt{\barp} Q$ and perform the related calculation. In the following, we will
go on like this. Thus, to cancel  the cross term, we set 
\begin{equation} \lb{condAoo}
\frac{\dot{\bar{\varphi}}^2}{c^2 \barp \bark} + 2 \kappa \frac{A_{00}}{\barp^2}  = 0 \leftrightarrow A_{00} = - \kappa \frac{\barp}{2} \frac{\dot{\bar{\varphi}}^2}{\bark}.
\end{equation}
With Eq. \ff{cond2}:
\begin{equation} \lb{condA01}
\frac{A_{01}}{a} = - \frac{1}{2 \bark} \left(\barp V' - \kappa \frac{\dot{\bar{\varphi}}^3}{2 \bark} \right).
\end{equation}
Finally, with the expressions of $A_{00}$, $A_{01}$, and $B_0$, one can see that the Hamiltonian is independent of $a$ and $b$, which means that the results is also independent of the
 normalization choice for the vectors  $A^m_{ai}$. One can also solve \ff{cond3} to find $A_{11}$ but this is physically unuseful. One can also study the
equations of evolution: once $\delta E$ is known, it is naturally possible to obtain the
 exact value of $\delta K$ by studying $\delta \dot{E}$. This leads to valuable informations on $\pi_a$.\\ 

Taking into account Eqs. \ff{condAoo} and \ff{condA01}, it is now possible to refine the expression of the Hamiltonian \ff{HGI1} as:
\begin{equation} \lb{Ham}
H^S_{GI} = \int \frac{d^3k}{2} \left[ \left(\frac{P}{\sqrt{\barp}}\right)^2  + \Gamma \left(\sqrt{\barp} Q\right)^2  \right],
\end{equation}
where $\Gamma$ is given by
\begin{equation}
\Gamma = k^2 + \barp V" + 3 \kappa \dot{\bar{\varphi}}^2 + 2 \kappa \frac{\dot{\bar{\varphi}}}{\bark} \barp V'- \frac{1}{2} \left(\kappa \frac{\dot{\bar{\varphi}}^2}{\bark} 
 \right)^2.
\end{equation}
To establish this expression, we have used  the Raychaudhuri equation (written in conformal time $\barN = \sqrt{\barp}$):
\begin{equation}
\dot{\mathcal{H}}=  \dot{\bark}= \bark^2 - \frac{\kappa}{2}  \dot{\bar{\varphi}}^2,
\end{equation}
and the Klein-Gordon equation:
\begin{equation}
\ddot{\bar{\varphi}}+  2 \bark  \dot{\bar{\varphi}} + \barp V' = 0.
\end{equation}

The Hamilton equations thus lead to:
\begin{eqnarray}
\dot{Q} &=& \frac{\partial H^S_{GI}}{\partial P} = \frac{P}{\barp}, \\
\dot{P} &=& - \frac{\partial H^S_{GI}}{\partial Q} = - \barp \Gamma Q,
\end{eqnarray}
and we recover the usual Mukhanov-like equation:
\begin{equation}
\ddot{Q} + 2 \bark \dot{Q} + \Gamma Q = 0,
\end{equation}
remembering that $ \bark = \frac{\dot{\barp}}{2 \barp}$. Classically, the Mukhanov equation 
is derived from the action:
\begin{equation}
\mathcal{S} = \int d\eta \int d^3k \frac{1}{2}\left[ \dot{v}^2 + \left(-  k^2 +  \frac{\ddot{z}}{z} \right) v^2 \right].
\end{equation}
The variable $v$ was here defined with $B_0=1$ such that
\begin{equation}
v = \sqrt{\barp} Q,
\end{equation}
leading, in the hamiltonian constraint, to:
\begin{equation}
\dot{v} = \bark v + \sqrt{\barp}\dot{Q} = \bark v + \frac{P}{\sqrt{\barp}} \rightarrow \frac{P}{\sqrt{\barp}} = \dot{v}- \bark v,
\end{equation}
and so:
\begin{eqnarray} 
H^{S}_{GI}&=& \int d^3 k \left[ \frac{1}{2} ( \dot{v}- \bark v)^2 + \frac{1}{2} \Gamma v^2
\right]. 
\end{eqnarray}
A standard Legendre transformation and an integration by part with respect to time leads to the Lagrangian: 
\begin{eqnarray}
\mathcal{L}^S  &=&\int d^3 k \left[ \frac{1}{2} \dot{v}^2 + \frac{1}{2} (-\Gamma +\bark^2 + \dot{\bark}) v^2  \right] \nonumber  \\
&=& \int d^3 k  \frac{1}{2} \left[ \dot{v}^2 - k^2 v^2  +  \frac{\ddot{z}}{z} v^2\right],
\end{eqnarray}
where terms leading to equivalent equations of motions have been deleted. With this equation, it is possible to recover
$z$ by solving:
\begin{equation} \lb{truerelation}
\frac{\ddot{z}}{z}  = - \Gamma + k^2  +\bark^2 + \dot{\bark},
\end{equation}
which is satisfied classically for 
\begin{equation}
z = \frac{\sqrt{\barp} \dot{\bar{\varphi}}}{\bark}.
\end{equation}

This method allowed us to find gauge-invariant variables and their equations of evolution, 
starting from an anomaly-free algebra  in an easy way. In Loop Quantum
Gravity, corrections to the classical theory are expected and this will change the expression for the constraints, leading to a modified algebra. 
The requirement of anomaly-freedom
is not necessary to obtain the gauge-invariant $Q$ and $P$, only a vanishing Poisson Bracket for the first order constraints is required.
But, of course, full physical consistency can only be achieved if the algebra is closed.
Thus, as in the previous sections where the case of General Relativity expressed with Ashtekar variables was studied, we will consider in the following the effects of 
the two main corrections from LQC, that is the holonomy and the inverse-volume corrections.

\section{Applications}

We now consider constraints modified with respect to the classical case. In the following, we focus only on the steps useful to find the associated gauge-invariant Mukhanov variables, without going into the details of the calculations. The missing steps can easily be rebuilt using the techniques given above.\\

We still consider a universe filled with a massive scalar field $\varphi$. The diffeomorphism constraint holds its classical form
and, in all the following considerations, it will still be given by Eq. \ff{diffeocons}. Moreover, the expressions of $\gamma_a$ and $\pi_a$ 
do not rely on the
shape of the constraints, but on the shape of the metric. In the following, expressions (\ref{gamma1solv2}-\ref{pimomenta2}) will therefore be used.
What will be modified are the Hamiltonian constraints where counter-terms have been added in order to cancel the anomalies so as to have a closed algebra. 
In the following, we will give the expressions of the first and second order for these constraints. The interested reader can go to the 
appendix where, starting from the zeroth
order constraints, the equations of motion for the background variables are derived. \\
To be as general as possible, we will keep the same notation where $a$ and $b$ are unknown. \\

\subsection{Inverse-volume case}

Following \cite{Bojowald:2008jv}, where anomaly-freedom was found in the case of 
inverse-volume corrections, we consider  hamiltonian densities given by:
\begin{eqnarray} 
&& \mathcal{H}^{(1)}_{G} = \frac{\bar{\alpha}}{2 \kappa} \left[ -4 (1+f)  \sqrt{\barp} \bark  \delta K^d_d \right. \nonumber \\
 && \left.- \frac{1}{\sqrt{\barp}}(1+g) \bark ^2 \delta E^d_d +  \frac{2}{\sqrt{\barp}}  \partial_c \partial^j \delta E^c_j \right],
\end{eqnarray}
\begin{eqnarray}
\mathcal{H}^{(1)}_{\pi} &=& \bar{\nu} \left[ (1+f_1)  \frac{\bar{\pi} \delta \pi}{\bar{p}^{3/2}}-(1+f_2)\frac{\bar{\pi}^2}{2\bar{p}^{3/2}} \frac{\delta^j_c \delta E^c_j}{2\bar{p}}  \right], \\
\mathcal{H}^{(1)}_{\varphi} &=& 
\bar{p}^{3/2} \left[ (1+f_3) V_{,\varphi}(\bar{\varphi}) \delta \varphi +V(\bar{\varphi}) \frac{\delta^j_c \delta E^c_j}{2\bar{p}} \right].
\end{eqnarray}
In this case, the Friedmann equation is:
\begin{equation} \lb{fried_IV}
\baralpha \bark^2 = \frac{\kappa}{3} \left(\frac{\barnu}{2} \frac{\barpi^2}{\barp^2} + \barp V \right),
\end{equation}
and, by definition,
\begin{equation}
\dot{\bar{\varphi}} = \barnu \frac{\barpi}{\barp}.
\end{equation}

Going ahead as in the classical case, Eq. \ff{Mnul} gives the relation between $B_0$ and $B_1$ such that:
\begin{eqnarray}
B_1 = - \frac{(1+f_1)}{(1+f)}\frac{ a }{ 2  \barp}\frac{  \dot{\bar{\varphi}} }{\baralpha \bark } B_0.
\end{eqnarray}
Setting $B_0 = 1$, the gauge-invariant Mukhanov-like variable is then:
\begin{eqnarray}
Q = \frac{\partial S}{\partial P} &=& \delta \varphi +  \frac{(1+f_1)}{(1+f)} \frac{\dot{\bar{\varphi}}}{\baralpha \bark} \psi.
\end{eqnarray}
Proceeding as before to solve the Hamilton-Jacobi-like equations, Eq. \ff{cond4} has to be fulfilled, which can be expressed here as:
\begin{equation} \lb{condxi2IV}
\xi_2 = \frac{b}{\kappa} \left[ - \baralpha \bark^2 (2f + g) + \frac{\kappa}{2} \frac{\dot{\bar{\varphi}}^2}{\bar{\nu}} (2 f_1 - f_2) \right].
\end{equation}
In our approach, this condition is satisfied if 
\begin{eqnarray}
g &=& - 2f, \\
f_2 &=& 2 f_1.
\end{eqnarray}
We have thus recovered exactly the relations given in \cite{Bojowald:2008jv} so as to have an anomaly-free algebra. This is of course  not surprising as
 \ff{condxi2IV} is related to the condition of anomaly-freedom that was pointed out in \ff{condAnomFree}.  \\
In \cite{Bojowald:2008jv}, the second order of the corrections $\alpha (\barp, \delta E^a_i)$ and $\nu (\barp, \delta E^a_i)$ also had to be taken into account,
 but as they are proportional to $\delta E$ and thus to $\gamma_1$ and $\gamma_2$, 
 we don't need to consider these terms. So, 
\begin{equation}
(\alpha^{(2)} ,\hskip 0.1truecm \nu^{(2)} ) = f (\gamma_1, \gamma_2)  \rightarrow  ignored.
\end{equation}
The second order constraint density with $\barN^a = 0$ can thus be written as:
\begin{eqnarray}
\mathcal{H}^{(2)} &=& \frac{\baralpha}{2 \kappa} \sqrt{\barp} ( \delta^c_k \delta^d_j \delta K^j_c \delta K^k_d  - (\delta K^d_d )^2 ) \nonumber \\
&&+  \frac{\barnu}{2} (1+g_1) \frac{\delta \pi^2}{\bar{p}^{3/2}} + \frac{\bar{\sigma}}{2} (1+g_5) \sqrt{\bar{p}} k^2 \delta \varphi^2 \nonumber \\
&& + \frac{1}{2} (1+g_6) \bar{p}^{3/2} V_{,\varphi\varphi}(\bar{\varphi}) \delta \varphi^2   + [\delta E^a_i].
\end{eqnarray}
In this case, the cross-term $PQ$ are vanishing if one imposes the following condition on $A_{00}$:
\begin{equation}
A_{00} = - \kappa \left( \frac{\dot{\bar{\varphi}}}{\barnu}\right)^2 \frac{\barp}{2 \bark}  \frac{(1+f_1)}{(1+f)} \frac{1}{(1+g_1)}.
\end{equation}
The Hamiltonian can therefore be written as:
\begin{equation}
H^S_{GI} = \int d^3 k \frac{1}{2} \left[\barnu (1+g_1) \left( \frac{P}{\sqrt{\barp}}\right)^2 +  \Gamma_{IV} (\sqrt{\barp} Q)^2 \right],
\end{equation}
where 
\begin{eqnarray}
&& \Gamma_{IV} = \kappa \left[  k^2 \bar{\sigma}(1+g_5) + \kappa (1+g_6) \barp V" + \frac{\dot{A}_{00}}{\barp} \right. \nonumber \\ 
&& \left. - 2 \baralpha \frac{A_{01}}{a}  \frac{\dot{\bar{\varphi}}}{\barnu}   + \frac{3}{2} \baralpha \left(\frac{\dot{\bar{\varphi}}}{\barnu} \right) ^2 + \barnu (1+g_1) \frac{A^2_{00}}{\barp^2} \right].
\end{eqnarray}
Let us define now
\begin{eqnarray} \lb{vIV}
v &=& \sqrt{\frac{\barp}{\barnu (1+g_1)}} Q, \\
\epsilon &=& \sqrt{\barnu (1+g_1)} \frac{d}{d\eta}\left( \frac{1}{\sqrt{\barnu (1+g_1)}}\right).
\end{eqnarray}
The Lagrangian can be written as 
\begin{equation}
\mathcal{L}^S_{GI} = \int \frac{d^3k}{2} \left( \dot{v}^2 +\left( -  k^2 \bar{\sigma}(1+g_5)  +  \frac{\ddot{z}}{z}\right) v^2 \right),
\end{equation}
where
\begin{eqnarray} \lb{eq_IV}
\frac{\ddot{z}}{z} &=& -\Gamma_{IV} +  k^2 \bar{\sigma}(1+g_5) + \left( \epsilon+ \frac{\baralpha \bark}{\sqrt{\barnu (1+g_1)}} \right)^2\nonumber \\
&&  + \frac{d}{d\eta} 
\left( \epsilon+ \frac{\baralpha \bark}{\sqrt{\barnu (1+g_1)}}\right).
\end{eqnarray}
Classically, the conserved curvature perturbation is given by:
\begin{equation}
R = \psi + \frac{\bark}{\dot{\bar{\varphi}}} \delta \varphi = \frac{v}{z}.
\end{equation}
In the case of inverse-volume corrections, from the previous equation and considering Eq. \ff{vIV}, one can naturally suggest in our approach:
\begin{equation}
z  =  \sqrt{\frac{\barp}{\barnu (1+g_1)}} \frac{(1+f_1)}{(1+f)} \frac{\dot{\bar{\varphi}}}{\baralpha \bark} ,
\end{equation}
which is close but not exactly similar to the expression given in \cite{arXiv:1011.2779}. In fact, the 
propagation speed for the perturbations derived in \cite{arXiv:1011.2779} is given by: 
\begin{equation}
s^2_{paper} = \baralpha^2 (1-f_3),
\end{equation}
whereas, in our case, it is equal to:
\begin{equation}
s^2 = \bar{\sigma}^2 (1+g_5).
\end{equation}
In this study, we have given some arguments to establish the function $z$. Although it would, in
principle, be possible to check its consistency, using Eq. \ff{eq_IV}, this would lead to lengthy
calculations that have not yet been carried out. It is however clear that our choice is associated with
a correct Lagrangian. It might be that both solutions are physically equivalent. We let this question
open for future studies.

\subsection{$\Omega$-LQC model -- holonomy corrections}

We now focus on the case of holonomy corrections and we use the notations  of \cite{arXiv:1111.3535}.

The first order corrected constraints, with counter-terms $\alpha_i$ not yet fixed but
introduced to close the algebra, are given by:
\begin{eqnarray}
&&\mathcal{H}^{(1)}_G = \frac{1}{2\kappa} \left( -4 \sqrt{\barp} (\K{s_1}+ \alpha_1 ) \delta K^d_d  \right. \nonumber \\
&& \left.- \frac{1}{\sqrt{\barp}} (\K{1}^2 +\alpha_2) \delta E^d_d + \frac{2}{\sqrt{\barp}}  \partial_c \partial^j \delta E^c_j \right),
\end{eqnarray}

\begin{eqnarray}
\mathcal{H}^{(1)}_{\pi} &=&  \frac{\bar{\pi} \delta
\pi}{\bar{p}^{3/2}}-\frac{\bar{\pi}^2}{2\bar{p}^{3/2}} \frac{\delta^j_c \delta E^c_j}{2\bar{p}},   \\
\mathcal{H}^{(1)}_{\varphi} &=& 
\bar{p}^{3/2} \left[ V_{,\varphi}(\bar{\varphi}) \delta \varphi +V(\bar{\varphi}) \frac{\delta^j_c \delta
E^c_j}{2\bar{p}} \right],
\end{eqnarray}

where we use the notation ($n \neq 0$):
\begin{equation}
\K{n} \hk \dot{=} \hk \frac{\sin (n \barmu \gamma \bark)}{n \barmu \gamma}.
\end{equation}
One also has to deal with the Klein-Gordon equation,
\begin{eqnarray}
\bar{\pi} &=& \barp \dot{\bar{\varphi}}, \\
\ddot{\bar{\varphi}} &=& - \barp \partial_{\bar{\varphi}} V(\bar{\varphi}) - 2 \K{2} \dot{\bar{\varphi}},  
\end{eqnarray}
and the Raychaudhuri equation,
\begin{equation}
\dot{\bark} = \bark \K{2} - \Omega \frac{\dot{\bar{\varphi}}^2}{2},
\end{equation}
with,
\begin{equation}
\Omega \dot{=} \cos (2 \barmu \gamma \bark).
\end{equation}
Moreover, holonomy corrections lead to a modified Friedmann equation. 
The energy density
$\rho$ is basically defined through:
\begin{equation} 
\K{1}^2 = \frac{\kappa}{3} \left(\frac{\dot{\bar{\varphi}}^2}{2}  + \barp V(\bar{\varphi}) \right) =
\frac{\kappa}{3} \barp \rho,
\end{equation}
and the Friedmann equation is given by:
\begin{equation}
\mathcal{H}^2 = \K{2}^2 = \frac{\kappa}{3} \barp \rho \left(1-\frac{\rho}{\rho_c}\right),
\end{equation}
where 
\begin{equation} \lb{rhoc}
\rho_c = \frac{3}{\kappa \gamma^2 \barmu^2 \barp }
\end{equation}
is the critical energy density. Applying the same procedure as previously, one can derive the relation between $B_0$ 
and $B_1$:
\begin{equation}
B_1 = - \frac{a \dot{\bar{\varphi}}}{2 \barp (\K{s_1} + \alpha_1)} B_0.
\end{equation} 
The related gauge-invariant variable is then (with the usual choice $B_0 = 1$):
\begin{equation}
Q = \delta \varphi +  \frac{\dot{\bar{\varphi}}}{ (\K{s_1} + \alpha_1)} \psi.
\end{equation}
In this case, the condition \ff{cond4} reads as:
\begin{equation}
0 = 2 \bark (\K{s_1} + \alpha_1) - 2 \K{1}^2 +\alpha_2,
\end{equation}
which is again a necessary condition appearing when $\{H_G^Q,D_G^Q\}$ is considered and which has to be 
fulfilled to have an anomaly-free algebra. By this procedure, we have
two unknown counter-terms $\alpha_1$ and $\alpha_2$, and the previous equation gives a relation between 
them:
\begin{equation}
\alpha_2 = 2 \K{1}^2 - 2 \bark (\K{s_1} + \alpha_1).
\end{equation}
In \cite{arXiv:1111.3535}, the anomaly was removed with $\alpha_1 = \K{2} - \K{s_1}$ and thus $\alpha_2 = 2 \K{1}^2 - 2 \bark \K{2}$ which will be used in the following.
The second order hamiltonian constraint density is thus modified such that:
\begin{eqnarray}
&&\mathcal{H}^{(2)} = \frac{\sqrt{\barp}}{2 \kappa} \cdot \Omega \cdot \left(  \delta^c_k \delta^d_j \delta K^j_c \delta K^k_d -  (\delta K^d_d )^2 \right) \nonumber \\
&&  + \frac{1}{2} \frac{\delta \pi^2}{\bar{p}^{3/2}} +  \frac{1}{2}  \bar{p}^{3/2} V_{,\varphi\varphi}(\bar{\varphi}) \delta \varphi^2  \nonumber \\
&&+  \frac{1}{2} \cdot\Omega \cdot \sqrt{\bar{p}} \delta^{ab} \partial_a \delta \varphi  \partial_b
\delta \varphi  + [\delta E].
\end{eqnarray}
One  obtains: 
\begin{equation} 
A_{00} = - \Omega \kappa \frac{\barp}{2} \frac{\dot{\bar{\varphi}}^2}{\K{2}},
\end{equation}
and:
\begin{equation} 
\frac{A_{01}}{a} = - \frac{1}{2 \K{2}} \left(\barp V' - \Omega \kappa \frac{\dot{\bar{\varphi}}^3}{2
\K{2}} \right),
\end{equation}

\begin{eqnarray}
\Gamma_\Omega &=& \Omega k^2 + \barp V" -\frac{\kappa}{2} \frac{\dot{\Omega}}{\K{2}} \dot{\bar{\varphi}}^2 + 3 \kappa \Omega \dot{\bar{\varphi}}^2 \nonumber \\
&& + 2 \kappa \Omega \barp V' \frac{\dot{\bar{\varphi}}}{\K{2}} - \frac{1}{2} \left(  \frac{\Omega \kappa
\dot{\bar{\varphi}}}{\K{2}} \right)^2.
\end{eqnarray}
The Mukhanov equation in conformal time ($\barN = \sqrt{\barp}$), remembering that 
$ \K{2} = \frac{\dot{\barp}}{2 \barp}$, is:
\begin{equation}
\ddot{Q} + 2 \K{2} \dot{Q} + \Gamma_\Omega Q = 0.
\end{equation}
As previously, it is possible to find $z$ through
\begin{equation} \lb{omegazz}
- \frac{\ddot{z}}{z}  = \Gamma_\Omega - \Omega \cdot k^2  +\K{2}^2 + \frac{d\K{2}}{d\eta},
\end{equation}
which is fulfilled for 
\begin{equation}
z = \frac{\sqrt{\barp}}{\K{2}} \dot{\bar{\varphi}}.
\end{equation}
This corresponds exactly to what was found, following another path based on
the Bardeen 
potentials, in \cite{arXiv:1111.3535}.
As in the previous case, we have recovered the  Mukhanov Lagrangian.

\subsection{General case: inverse-volume and holonomy corrections}
In this section, we will not address the issue of  anomaly-freedom for the case where both corrections 
are taken into account. We will just focus on defining the Mukhanov variable by the
method  previously developed. Naturally, it then will be expressed as a function of counter-terms. 
We will see that this case can be solved as if  corrections were mostly independent, as suggested in 
\cite{arXiv:1011.2779}.   \\
In this case, the first order constraint densities can be defined, as in the previous case, such that:

\begin{eqnarray}
\mathcal{H}_G^{(1)} &=& \frac{\baralpha}{2 \kappa} \left[ - 4\sqrt{\barp}(\K{s_1} + \alpha_1) \delta K^d_d \right.
\nonumber \\
&& \left.- \frac{1}{\sqrt{\barp}} (\K{1}^2 + \alpha_2) \delta E^d_d   +\frac{2}{\barp}   \partial_c
\partial_j \delta E^c_j \right],  ~~~
\end{eqnarray}
 for the gravitationnal sector and,
\begin{eqnarray}
\mathcal{H}^{(1)}_{\pi} &=&  \barnu (1+f_1) \frac{\bar{\pi} \delta \pi}{\bar{p}^{3/2}}- \barnu (1+f_2)
\frac{\bar{\pi}^2}{2\bar{p}^{3/2}} \frac{\delta^j_c \delta E^c_j}{2\bar{p}},   \\
\mathcal{H}^{(1)}_{\varphi} &=& 
\bar{p}^{3/2} \left[(1+f_3)  V_{,\varphi}(\bar{\varphi}) \delta \varphi +V(\bar{\varphi})
\frac{\delta^j_c \delta E^c_j}{2\bar{p}} \right],~~~
\end{eqnarray}
for the matter sector.

The Friedmann-like equation is given by:
\begin{equation} \lb{fried_IVholo}
\baralpha \K{1}^2 = \frac{\kappa}{3} \left(\frac{\barnu}{2} \frac{\barpi^2}{\barp^2} + \barp V \right),
\end{equation}
with the definition:
\begin{equation}
\dot{\bar{\varphi}} = \barnu \frac{\barpi}{\barp}.
\end{equation}
Going ahead as previously, one obtains again a relation between $B_0$ and $B_1$:
\begin{eqnarray}
B_1 = - \frac{(1+f_1)}{(1+f)}\frac{ a }{ 2  \barp}\frac{  \dot{\bar{\varphi}} }{\baralpha (\K{s_1} +
\alpha_1) } B_0.
\end{eqnarray}
As a fully closed algebra has not yet been found, we are compelled to keep $\alpha_1$ as unknown. Using
previous results from \cite{arXiv:1011.2779} and \cite{arXiv:1111.3535}, it is however probable that the solution
will soon be derived. 

Setting $B_0 = 1$, the gauge-invariant Mukhanov-like variable is then:
\begin{eqnarray}
Q = \frac{\partial S}{\partial P} &=& \delta \varphi +  \frac{(1+f_1)}{(1+f)}
\frac{\dot{\bar{\varphi}}}{\baralpha (\K{s_1} + \alpha_1) } \psi.
\end{eqnarray}

Following the same procedure as in the previous sections when solving the Hamilton-Jacobi-like equations,
we  have to fulfill, in particular, Eq. \ff{cond4}, which can be expressed here as:
\begin{eqnarray} 
\xi_2 = \frac{b}{\kappa} &&\left[  - \baralpha (2 \bark (\K{s_1} + \alpha_1) + \alpha_2 - 2 \K{1}^2) \right. \\
&&\left.  +  \frac{\kappa}{2} \frac{\dot{\bar{\varphi}}^2}{\bar{\nu}} (2 f_1 - f_2) \right], 
\end{eqnarray} 
and vanishes for 
\begin{eqnarray}
2 \bark (\K{s_1} + \alpha_1) &+& \alpha_2 - 2 \K{1}^2 = 0,\\
f_2 &=& 2 f_1.
\end{eqnarray}
This corresponds to conditions already found when holonomy or inverse-volume corrections were
taken into account independently.

At this stage, it is difficult to go much ahead. However, the previous results 
lead us to assume 
\begin{equation}
z  =  \sqrt{\frac{\barp}{\barnu (1+g_1)}} \frac{(1+f_1)}{(1+f)} \frac{\dot{\bar{\varphi}}}{\baralpha (\K{s_1} + \alpha_1)}.
\end{equation}
This remains to be fully demonstrated.

\section{conclusion}

This article builds on the innovative ideas given in \cite{Langlois} and develop them so that they can be used
in the framework of Loop Quantum Cosmology. Going through successive changes of variables by using
appropriate generating functions, we have set a scheme useful for the definition of gauge-invariant
variables appropriate to study cosmology. Both the cases of inverse-volume and of
 holonomy corrections were considered. 
In principle, it is also possible to study in such an easy way, any other kind of 
correction that can be set up by contraints.
Moreover, the Hamilton-Jacobi method is very  general and can be used 
in different frameworks, such as particle physics, 
where gauge-invariance is  required.\\

Several developments are expected. First, although reasonable, some guesses had to be made. They
should be checked in details by going though the full exhaustive computation, 
in particular by studying the equations of motion for the Bardeen potentials.

Second, the method should be applied again when counter-terms for the holonomy + inverse-volume case
will have been found. This work is already on the way.

Finally, those gauge-invariant variables should now be used to investigate cosmological consequences
at the observational level.

\section{appendix : equations of motion }
In this appendix, we give an example of how to derive the equations of motion for the background 
variables when holonomy and inverse-volume corrections are taken into account together. The classical
limit corresponds to $\baralpha \rightarrow 0$ or $\K{n} \rightarrow \bark$. 
At the lowest order, the diffeormorphism constraints are null, we therefore consider
in the following only the hamiltonian constraints which are given, for  gravity and for matter,
by: 
\begin{equation}
H_G^{(0)}[\barn] = \frac{1}{2 \kappa} \int d^3x \barN \left(  - 6 \baralpha \K{1}^2 \sqrt{\barp}  \right),
\end{equation}
and 
\begin{equation}
H_M^{(0)}[\barn] =  \int d^3x \bar{N} \left( \barnu \frac{\bar{\pi}^2}{2\bar{p}^{3/2}}+ \bar{p}^{3/2}
V(\bar{\varphi})\right).
\end{equation}
The background variables are linked through:
\begin{eqnarray}
\{\bark, \barp \} &=& \frac{\kappa}{3}, \\
\{\bar{\varphi}, \barpi \} &=& 1.
\end{eqnarray}
For instance, the conformal Hubble parameter found with $\barp = a^2(\eta)$ is given by:
\begin{eqnarray}
\mathcal{H} = \frac{\dot{\barp}}{2 \barp} &=&  \frac{1}{2 \barp} \{\barp, H_G^{(0)}[\barn]+  H_M^{(0)}[\barn] \} \nonumber \\
&=& \frac{1}{2 \barp}  \frac{\kappa}{3 } \left[ \frac{\partial \cdot}{\partial \bark} \frac{\partial \cdot}{\partial \barp} - \frac{\partial \cdot}{\partial \barp} \frac{\partial
\cdot}{\partial \bark} \right] \\
&=& 0 - \frac{1}{2 \barp}  \frac{\kappa}{3} \frac{\partial \barp }{\partial \barp} \left( \frac{\partial H_G^{(0)}[\barn]}{\partial \bark}  +0  \right)    \\
&=& \frac{1}{2 \barp}  \frac{\kappa}{3} \frac{\sqrt{\barp}}{2 \kappa} \left( - 12 \sqrt{\barp} \baralpha \K{2}\right)\\
\mathcal{H} = \frac{\dot{\barp}}{2 \barp} &=&  \baralpha \K{2}.
\end{eqnarray}
Moreover, the energy density $\rho$ is defined by 
\begin{equation}
\rho  \dot{=}  \frac{1}{\barp^\frac{3}{2}} \frac{\delta H_m}{\delta \barN} = \frac{\bar{\nu}}{2}
\frac{\barpi^2}{\barp^3} + V,
\end{equation}
and is linked to the gravity through an equation of motion
\begin{equation}
\frac{\delta}{\delta  \bar{N} } (H+D)^{(0)}_{tot} = 0,
\end{equation}
which gives the Friedmann-like equation
\begin{equation} \lb{preFried}
\baralpha \K{1}^2 = \frac{\kappa}{3} \left(\frac{\barnu}{2} \frac{\barpi^2}{\barp^2} + \barp V \right) =
\frac{\kappa}{3} \barp \rho.
\end{equation}
Using the previous relations, the Friedmann equation is thus given by:
\begin{eqnarray}
\mathcal{H}^2 &=& (\baralpha \K{2})^2  = \baralpha (\baralpha \K{1}^2)\cdot cos^2(\barmu \gamma \bark)  \\
&=&  \baralpha (\baralpha \K{1}^2) (1 - (\barmu \gamma)^2 \K{1}^2 )  \\
&=& \baralpha \left(\frac{\kappa}{3} \barp \rho \right) \left( 1 - \frac{1}{\baralpha} \frac{\rho}{\rho_c} \right) \\
&=&  \frac{\kappa}{3} \barp \rho \left( \baralpha -  \frac{\rho}{\rho_c} \right),
\end{eqnarray}
where, as usual when using the holonomies within the $\barmu$-scheme, the critical energy density is defined by Eq. \ff{rhoc}.
Moreover, similarly, the Raychaudhuri equation is given by
\begin{equation}
\dot{\bark} = \baralpha \bark \K{2} - \left(\frac{\barp}{\baralpha} \frac{\partial \baralpha}{\partial \barp} \right)\baralpha \K{1}^2 - \frac{\kappa}{2}
\frac{\dot{\bar{\varphi}}^2}{\barnu}.
\end{equation}

\[\]\[\]\[\]

\end{document}